\documentstyle[a4,12pt,epsf]{article}
\newcommand{\noi}{\noindent}
\newcommand{\eq}{\begin{equation}}
\newcommand{\en}{\end{equation}}
\newcommand{\eqa}{\begin{eqnarray}}
\newcommand{\ena}{\end{eqnarray}}
\newcommand{\eqann}{\begin{eqnarray*}}
\newcommand{\enann}{\end{eqnarray*}}

\newcommand{\pbp}{\bar{\psi}\psi}

\newcommand{\tr}{\mbox{Tr}\,}
\newcommand{\cam}{{\cal M}}

\newcommand{\aleq}{\mbox{}_{\textstyle \sim}^{\textstyle < }}
\newcommand{\ageq}{\mbox{}_{\textstyle \sim}^{\textstyle > }}
\newcommand{\ra}{\rightarrow}

\newcommand{\be}{\begin{equation}}
\newcommand{\ee}{\end{equation}}
\newcommand{\bea}{\begin{eqnarray}}
\newcommand{\eea}{\end{eqnarray}}

\newcommand{\bt}{\beta}

\begin{document}

\hbox{ }
\noi {\small June 1996} \hfill {HU Berlin--EP--96/17}   \\
\mbox{} \hfill IFUP--TH 29/96                              \\
\mbox{} \hfill SWAT/96/108                                  \\

\renewcommand{\thefootnote}{\fnsymbol{footnote}}

\begin{center}

{\LARGE Efficiency of different matrix inversion
methods applied to Wilson fermions}
\footnote{ Work partly supported by EEC--Contract CHRX--CT92--0051}

\vspace*{1.0cm}
{\large
G.~Cella$\mbox{}^1$,
A.~Hoferichter$\mbox{}^2$,
V.K.~Mitrjushkin$\mbox{}^{2,3}$
\footnote{Permanent address: Joint Institute for Nuclear Research, 
Dubna, Russia},
M.~M\"uller--Preussker $\mbox{}^2$
and
A.~Vicere$\mbox{}^1$
}\\
\end{center}
\vspace*{0.5cm}
{\normalsize
$\mbox{}^1$ {\em INFN in Pisa and Dipartimento di Fisica dell'Universit\'a
di Pisa,  Italy }\\
$\mbox{}^2$ {\em Humboldt-Universit\"{a}t zu Berlin, Institut f\"{u}r Physik,
10099 Berlin, Germany}\\
$\mbox{}^3$ {\em Department of Physics, University of Wales, Swansea, U.K. }\\
}  %end normalsize

\vspace*{1cm}
\begin{center}
{\bf Abstract}
%================================================================
\end{center}

\noi We compare different conjugate gradient -- like matrix 
inversion methods (CG, BiCGstab1 and BiCGstab2) employing for
this purpose the compact lattice quantum electrodynamics 
(QED) with Wilson fermions.
The main goals of this investigation are the CPU time efficiency
of the methods as well as the influence of machine precision on 
the reliability of (physical) results especially 
close to the 'critical' line $~\kappa_c(\bt)$.

\renewcommand{\thefootnote}{\arabic{footnote}}
\setcounter{footnote}{0}

\section{Introduction}
%================================================================

In many physical applications it is necessary to perform the 
inversion of some  large $~N \times N~$ matrix
$~{\cal M}$, i.e., to solve the equation

\eq
{\cal M} \cdot X =\varphi~,
                       \label{Mx=p}
\en

\noi where $~\varphi~$ is some known input--vector and $~X~$ the
required solution. It becomes not an easy computational task
to calculate the vector $~X~$ when, for example,  $~N \ageq 10^4~$ 
especially when $~{\cal M}~$ is not well--conditioned.
The problem of {\it efficiency} and {\it reliability} 
of the inversion algorithm appears to be of crucial importance.
The lattice approach to gauge theories (e.g., QED or QCD) 
which provides a powerful tool for the numerical 
study of quantum field theories in the nonperturbative regime
is a typical example.
In the  case of the $~U(1)~$ gauge group and Wilson fermions (QED) 
the matrix to be inverted is  \cite{wil}

\eqa
{\cal M}(U;\kappa ) & \equiv & \hat{1} - \kappa \cdot Q(U) ,
\nonumber \\ \nonumber \\
Q_{xy}(U)
& = & \sum_{\mu =1}^4
\Bigl[ \delta_{y, x+\hat{\mu}} \cdot ( {\hat 1} - \gamma_{\mu})
\cdot U_{x \mu} +
\delta_{y, x-\hat{\mu}} \cdot ( {\hat 1} + \gamma_{\mu})\cdot
U_{x-\hat{\mu}, \mu}^{\dagger} \Bigr]~,
                                              \label{waf}
\ena

\noi where $~x,y~$  denote the sites of a $~4d~$ lattice,
$~{\hat \mu}~$ is the unit vector in the direction $~\mu~$,
and $~\gamma_{\mu}$'s are $~4\times 4~$ Dirac's matrices
in euclidean space.
The set of gauge field link variables 
$~\left\{ U_{x\mu} \right\},~~ U_{x\mu} \in U(1) ~$
generated with the proper statistical weight defines
the corresponding gauge field configuration, and the hopping
parameter $~\kappa~$ is related to the bare mass of the fermion \cite{wil}.
The dimension of the fermionic 
matrix is $~N\times N$ with $N=4 \cdot N_{sites}$,
and the typical number of sites in lattice calculations is
$~N_{sites} \sim 8^4 \div 20^4~$.

In our study 
we were especially interested in the physically most interesting 
-- and technically most difficult -- 'extreme' case,  when the 
minimal eigenvalue $~\lambda_{min}~$ of the fermionic matrix $~{\cal M}~$
tends to zero. 
In the case of Wilson's fermionic matrix this limit is
connected with the special choice of the hopping parameter
$~\kappa \ra \kappa_c(\beta )$, $~\beta~$ being the gauge 
coupling parameter, and 
is supposed to correspond to 
the chiral limit of the theory (see, e.g., \cite{hmmns}).
By approaching the chiral limit within the confinement
phase the most common matrix inversion methods like 
conjugate gradient (CG) or minimal residue (MR) become extremely 
inefficient or fail (for review see, e.g., \cite{templ})
according to the fact that the condition number of the fermionic matrix
diverges.

Recently new very promising cg--like matrix inversion algorithms have 
been proposed : BiCGstab1 \cite{bcg1a} and BiCGstab2 \cite{bcg2a}. 
The successful application of BiCGstab1 to lattice gauge theory issues 
has been demonstrated in \cite{bcg1b} for intermediate quark masses,
i.e., for $~\kappa$'s  sufficiently below $~\kappa_c(\beta)$.
In this case  the CPU time improvement factor was reported to be
of about 2$\div$3 compared to CG. 
Similar improvement was reported also for the BiCGstab2 method 
\cite{bcg2b,prog_qcd}.
However, the problem of the efficiency and reliability of these
methods in the $~\lambda_{min} \to 0~$ limit deserves further study.
A delicate point to be studied is the possible breakdown 
because of the accumulation of roundoff errors.
This accumulation
becomes much more dangerous when approaching the chiral limit and/or
when the precision of the used machine is not 
large, as in the case of 
APE/Quadrics. On this machine, the
native single precision
\footnote{i.e., 32bit precision}
arithmetic can be improved using accurate summation methods
\cite{lues_ks} but which cannot fully compete with hardware supported
double precision
and require more computational power.
Another interesting question which did not attract still sufficient
attention is the the connection between the convergence of the
residue vectors and the convergence of observables.
As it will be shown, this connection can be rather nontrivial.

The above mentioned topics constitute the main goals of this work.
In the second section we give a short description 
of the inversion methods under study and main notations.
The third section is devoted to the reliability of the inversion 
methods, while in section four we present the CPU time 
improvement factors. The last section contains our conclusions.

\section{Methods and observables}

The structure of the fermionic matrix $~{\cal M}~$ in eq.(\ref{waf})
permits 
an optimization based on an even--odd
decomposition \cite{degr1,degr2} :
\eq
{\cal M} = \left(
\begin{array}{cc}
\hat{1}_o & \kappa \cdot Q_{oe} \\
\kappa\cdot  Q_{eo} & \hat{1}_e \\
\end{array}
\right),
                             \label{eo}
\en
\noi where subscripts {\it e} and {\it o} stand for the even and odd 
subspace, correspondingly. 
Therefore, it is sufficient to invert, say, the matrix $~{\cal D}_e~$
\eq
{\cal D}_e =\hat{1}_e -\kappa^2 Q_{eo}Q_{oe}~,
\en
\noi defined only on the even subspace in order 
to solve the problem (\ref{Mx=p}) since the solution vector 
of the complementary subspace can be easily constructed from 
the solution $~X_e$. Throughout this work we refer to the 
even--odd preconditioned versions of the investigated algorithms. 

We made a comparative study of the standard conjugate 
gradient method (CG), biconjugate gradient stabilized (BiCGstab1)
and  biconjugate gradient stabilized II  (BiCGstab2).
These methods approximate the
true solution iteratively
by generating a sequence $~X_0,X_1,\ldots ,X_k,\ldots $, where
$~X_0~$ denotes the starting vector which
can be chosen with different methods, for example in a
random way.
For completeness we give a short schematic description of these  methods.
To keep the notation simple we drop the subspace 
index in the algorithmic part of the formulas.

\begin{itemize}

\item {\bf CG}. The equation to solve 
is ${\cal D} X_e = \varphi^{\prime\prime}_e=
{\cal D}_e^{\dagger} \varphi^{\prime}_e = 
{\cal D}_e^{\dagger}\bigl(\varphi_e - \kappa Q_{eo}\varphi_o \bigr)$, with
${\cal D} = {\cal D}_e^{\dagger} {\cal D}_e$.

$P_0  =  R_0 = \varphi^{\prime\prime} - {\cal D} X_0 \\$
For ~$k  =  0, \ldots ~$:\hfill
\vspace{-1.4cm}
\eqa
\nonumber \\
\nonumber \\
X_{k+1} & = & X_{k} + \omega_{k}  P_{k}~;
\quad
\omega_k =\bigl( R_{k},R_{k} \bigr) / \bigl( P_{k},{\cal D}P_{k} \bigr)
\nonumber \\
R_{k+1} &=& R_{k} - \omega_{k}  {\cal D} P_{k}
\nonumber \\
P_{k+1} &=& R_{k+1} + \gamma_{k+1}  P_{k}~;
\quad  \gamma_{k+1} =
\bigl( R_{k+1},R_{k+1} \bigr) / \bigl( R_{k},R_{k} \bigr)~.\qquad\qquad
\nonumber
\ena

\noi This method is simple in implementation and reliable,
though it is not the fastest. 
In the version with double precision
we use it as a reference point
for the other two methods.

\item {\bf BiCGstab1}. The equation to solve is 
${\cal D} X_e = \varphi^{\prime}_e$, with ${\cal D} = {\cal D}_e$.

$ V_0 = P_0 = 0 \\
\alpha_0 = \omega_0 = \rho_0 = 1 \\
R_0 = \hat{R}_0 = \varphi^{\prime} - {\cal D}X_0 $\\ \\
For ~~$k  =  1, \ldots ~$:\hfill
\vspace{-0.2cm}
\eqann
\rho_k &=&  \bigl(\hat{R_0},R_{k-1}\bigr) ~;
\quad 
\beta_k =  \bigl ( \rho_k  \alpha_{k-1}\bigr ) \big /
\bigl (\rho_{k-1}  \omega_{k-1}\bigr ) \qquad\qquad\qquad\qquad~~~
\\
P_{k} &=&  R_{k-1} + \beta_k \bigl( P_{k-1}-\omega_{k-1}V_{k-1} \bigr)
\qquad\qquad\qquad\qquad\qquad\qquad
\\
V_k&=&{\cal D}P_k ~;
\quad
\alpha_k = \rho_k \big / \bigl(\hat{R}_0,V_k\bigr)
\\
S_k&=&R_{k-1}-\alpha_k V_k
\\
T_k&=&{\cal D}S_k ~; 
\quad
\omega_k = \bigl(T_k,S_k\bigr)\big /\bigl(T_k,T_k\bigr)
\\
X_k&=&X_{k-1}+\omega_k S_k + \alpha_k P_k
\\
R_k&=&S_k - \omega_k T_k~.
\\
\enann
\vspace{-1.5cm}

\item {\bf BiCGstab2}. The equation to solve is 
${\cal D} X_e = \varphi^{\prime}_e$, with ${\cal D} = {\cal D}_e$.

%%%%%%%%%%%%%%%%%%%%%%%%%%%%%%%%%%%%%%%%%%%%%%%%%%%%
% 18/03/96                                         %
% File: bicgs2.tex                                 %
%%%%%%%%%%%%%%%%%%%%%%%%%%%%%%%%%%%%%%%%%%%%%%%%%%%%

$ R_0 = \hat{R}_0 = \varphi^{\prime} - {\cal D}X_0 \\
\mbox{choose}~~ Y_0~~\mbox{such that}~~
\delta_0=\bigl(Y_0,R_0\bigr) \not= 0 ~~
\mbox{and}~~\omega_0^{-1}=\bigl(Y_0,{\cal D}\hat{R}_0\bigr)/\delta_0
\not= 0  \\$
For $k  =  0, \ldots ~ :$
\vspace{-0.2cm}
\eqann
V_{k+1} &=& U_k - \omega_k {\cal D}T_k ~,~~k \geq 1 \qquad\\
U_{k+1} &=& R_k - \omega_k {\cal D}\hat{R}_k \\
\lefteqn{\mbox{\hspace{-1.5cm}\underline{If}}~~k~~
\mbox{even~~\underline{then}}~~m:=k/2} \\
\chi_m &=& \bigl({\cal D}U_{k+1},U_{k+1}\bigr)\big / 
\bigl({\cal D}U_{k+1},{\cal D}U_{k+1}\bigr)
\\
R_{k+1} &=& U_{k+1} - \chi_m {\cal D}U_{k+1} \\
X_{k+1} &=& X_k + \omega_k\hat{R}_k +\chi_m U_{k+1} \\
\delta_{k+1} &=& \bigl(Y_0,R_{k+1}\bigr)~;
\quad 
\psi_{k+1} = - \bigl(\omega_k \delta_{k+1}\bigr)\big/\bigl(\chi_m\delta_k\bigr) \\
\hat{R}_{k+1} &=& R_{k+1} - \psi_{k+1}(\hat{R}_k 
- \chi_m{\cal D}\hat{R}_k)\\
\lefteqn{\mbox{\hspace{-1.5cm}\underline{else}}~~~ m:=(k-1)/2}\\
B_{m+1} &=& [U_{k+1}-V_{k+1} | {\cal D}U_{k+1}] 
~~\mbox{(2 column matrix)}\\
\left( 
\begin{array}{c}
\xi_m \\
\eta_m\\
\end{array}
\right) &=& -(B_{m+1}^{+}B_{m+1})^{-1}B_{m+1}^{+}V_{k+1} \\
R_{k+1} &=& (1-\xi_m)V_{k+1} + \xi_m U_{k+1} 
+ \eta_m {\cal D} U_{k+1} \\
X_{k+1} &=&
(1-\xi_m)(X_{k-1} + \omega_{k-1}\hat{R}_{k-1} + \omega_k T_k) 
+ \xi_m(X_k +\omega_k\hat{R}_k) - \eta_m U_{k+1} \\
\delta_{k+1} &=& \bigl(Y_0,R_{k+1}\bigr)~;
\quad
\psi_{k+1} = \bigl(\omega_k \delta_{k+1}\bigr)\big/\bigl(\eta_m \delta_{k}\bigr)\\
\hat{R}_{k+1} &=& R_{k+1} -\psi_{k+1} \Bigl ((1-\xi_m)T_k + \xi_m\hat{R}_k
+ \eta_m{\cal D}\hat{R}_k \Bigr ) \\
\lefteqn{\mbox{\hspace{-1.5cm}\underline{endif}}}\\
T_{k+1} &=& U_{k+1} - \psi_{k+1}\hat{R}_k~; \quad
{\cal D}T_{k+1} = {\cal D} U_{k+1} - \psi_{k+1}{\cal D}\hat{R}_k \\
\omega_{k+1} &=& \delta_{k+1}\big /\bigl(Y_0,{\cal D}\hat{R}_{k+1}\bigr)~.\\
\enann
\end{itemize}

\noi Compared to BiCGstab1 this algorithm has more complicated 
recurrences and requires more dot product operations. 

\vspace{0.2cm}

The iterative procedure 
was 
stopped when $~||R_k|| < \varepsilon~$. We used
$~\varepsilon=10^{-5}~$ when implemented on an APE/Quadrics system
(equivalent to the condition
$||R_k||_{norm}\equiv||R_k||/||R_0||~ \aleq~ 10^{-7}~$ on an
$~8^3\times 16~$ lattice), and  $~\varepsilon=5\cdot 10^{-8}~$
on double precision machines. 
A point to mention is that the matrix to be inverted
in the case of CG  is ${\cal D} = {\cal D}_e^{\dagger} {\cal D}_e$
and hence differs from the BiCGstab1/2 cases where ${\cal D} = {\cal D}_e$. 
The application of the same stopping criterion $~||R_k|| < \varepsilon~$
to the different algorithms entails a systematic error, which we have found
to be 0$\div$2\%. 

\noi The algorithmic residue vectors $~R_k^{algo} = R_k~$ can differ 
substantially from the corresponding true residue vectors 
$~ R_k^{true} = \varphi - {\cal D} X_k~$  due to roundoff errors.
Apart from the norms of the two residue vectors ($||R^{true}_k||~$ and 
$||R^{algo}_k||$) we monitored also for every configuration  the convergence of 
the pion norm $~\Pi~$, scalar condensate 
$~{\bar \psi}\psi~$ and pseudoscalar condensate 
$~{\bar \psi} \gamma_5 \psi$ defined as :

\eqa
\Pi 
& = & \frac{1}{4N_{sites}} \cdot \mbox{Tr} \Bigl[ {\cal M}^{-1}
\gamma_{5} {\cal M}^{-1} \gamma_{5} \Bigr] ~;
                                      \label{pionnm}
\\ \nonumber \\
{\bar \psi} \psi & = & \frac{1}{4N_{sites}} \cdot
 \mbox{Tr} \Bigl[{\cal M}^{-1} \Bigr] ~;
\qquad
{\bar \psi} \gamma_{5}  \psi =\frac{1}{4N_{sites}} \cdot
\mbox{Tr} \Bigl[ \gamma_{5} {\cal M}^{-1}\Bigr] ~. \label{pg5p}
\ena

\noi It is useful to write down the spectral representation
of these observables

\eqa
\Pi &=& \frac{1}{4N_{sites}} \sum_{n} \frac{1}{\mu_{n}^2}~;
\nonumber \\ \nonumber \\
{\bar \psi} \psi &=& \frac{1}{4N_{sites}} \sum_{n} \frac{1}{\lambda_{n}}~;
\qquad
{\bar \psi} \gamma_{5} \psi =\frac{1}{4N_{sites}} \sum_{n} \frac{1}{\mu_{n}}~,
                                 \label{spectral}
\ena

\noi where $~\lambda_n~$ and $~\mu_n~$ denote the eigenvalues 
of $~{\cal M}~$ and $~{\cal M}\gamma_5 $, respectively.
Since $~\mu_n \to 0~$ when the corresponding $~\lambda_n \to 0$, it becomes
clear why $~\Pi~$ is a very sensible observable near $~\kappa_c(\bt)$.

\noi The conventional definition of the bare fermion mass $~m_q~$ is

\eq
m_q = \frac{1}{2}\cdot\Bigl(\frac{1}{\kappa} - 
\frac{1}{\kappa_c(\beta)}\Bigr).
						\label{mq}
\en

To monitor the influence of the accuracy we performed
our calculations on a Cray--YMP 
and Siemens--Fujitsu (default 64bit, i.e., double precision),
Convex--C3820--ES (single and double precision) and a
QH2  -- $~8\times 8\times 4~$ APE/Quadrics semitower.
The native single precision of QH2 was improved by implementation
of the Kahan summation method 
applied to global sums \cite{lues_ks}. 

The lattice sizes we used have been $~16^3\times 32~$ and 
$~32^3\times 64~$ on the QH2, and $~8^3\times 16~$ and $~16^3\times 32~$
on the vector machines.
The choice of the gauge coupling parameter $~\beta~$ was $~0~$ and $~0.8~$
in the confinement phase and $~1.1~$ in the Coulomb phase.

\section{Reliability and precision effects}
%=================================================

\subsection{CG in single and double precision modes}

\begin{figure}[p]
\vspace{-3.0cm}
\epsfysize=570pt\epsfbox{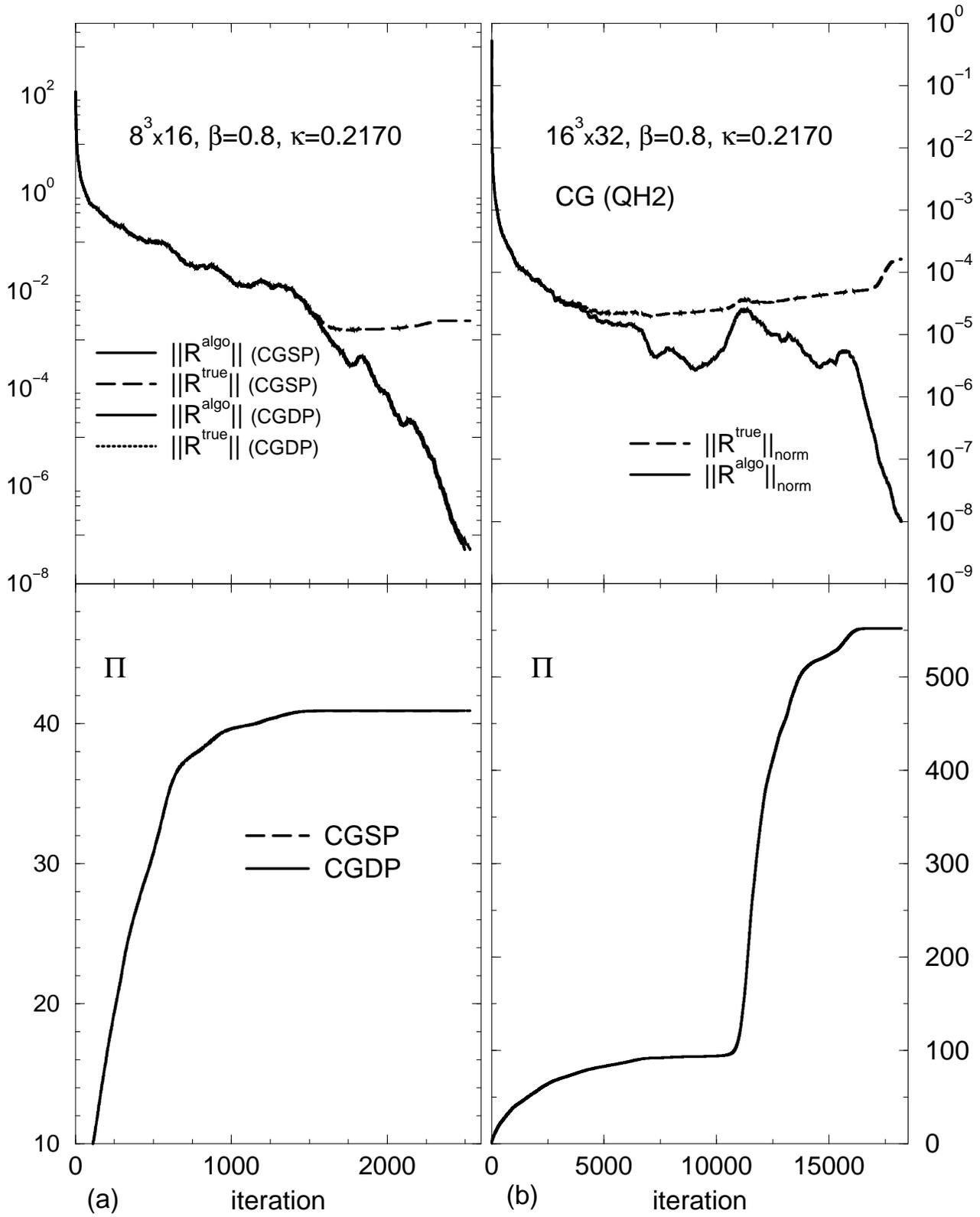}
\vspace{1.0cm}
\caption{
Convergence histories on single configurations on a vector machine (a) 
and QH2 (b).
$||R_k||_{norm} = ||R_k||/||R_0||$.
}
\label{fig:R_CX_SP_QH2}
\end{figure}

One way to study the reliability of the inversion method
is to monitor the inversion history of the two residues :
$~R^{true}~$ and $~R^{algo}$.
In the case of the double precision conjugate gradient method (CGDP) 
the norms of these two residues coincide with high accuracy down 
to very small values which ensures the reliability of the result.
In Figure \ref{fig:R_CX_SP_QH2}a we show an example of such a run 
at $~\beta=0.8~$ and $~\kappa =0.2170~$ on an $~8^3 \times 16~$ lattice.
The value of the hopping parameter $~\kappa~$ was chosen
to be rather close to the critical value 
$~\kappa_c(\beta =0.8) \simeq 0.2171(1)$.

To our experience, CGDP never fails, although can become very slow.
Therefore, we used the double precision 
conjugate gradient method  as a reference point for 
the two 'stabilized' methods and for CG in single precision mode (CGSP). 

To single out the effect of machine accuracy, we applied the
single precision and double precision CG to the {\it same}
configurations providing the same startvectors and sourcevectors
(see, e.g., Figure \ref{fig:R_CX_SP_QH2}a).
%
%  fig_3_1a
%
The single precision residue $~|| R_k^{true}||~$ decouples 
from the other residues after $~\sim 1500~$ iterations 
due to lack of accuracy.
However, the single precision algorithmic residue $~||R_k^{algo}||~$
coincides with both double precision residues.  In this case the pion
norm $~\Pi~$ practically does not depend on the machine accuracy.  It
reaches its plateau before $~||R_k^{true}||~$ has decoupled from the other
residues.

%
%
%
%  fig_3_1b
%
\begin{figure}[p]
\vspace{-3.0cm}
\epsfysize=570pt\epsfbox{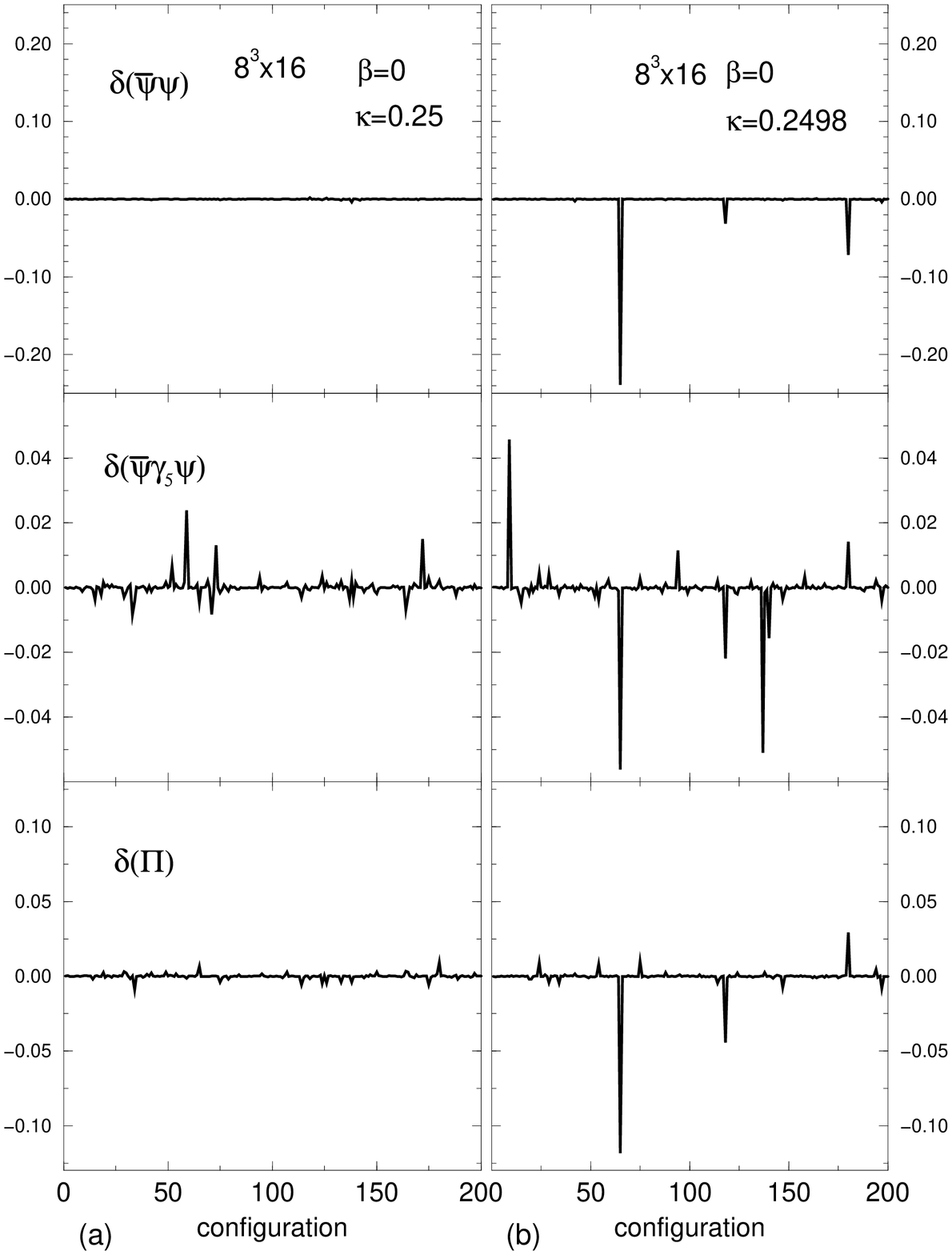}
\vspace{1.0cm}
\caption{
The relative deviation $~\delta~$ as defined in eq.(\protect\ref{dev})
at $~\kappa=0.25$ (a) and $~\kappa=0.2498$ (b).
}
\label{fig:3_1b}
\end{figure}

In some cases, however, the decoupling of $~||R_k^{algo}||~$ and
$~||R_k^{true}||~$ occurs before the observable (e.g., the pion norm)
reaches its plateau.  An example of such an inversion history on the QH2
(single precision)
for $~\kappa =0.2170~$ is shown in Figure {\ref{fig:R_CX_SP_QH2}}b.
Despite the Kahan improved summation, $~||R^{true}_k||~$ starts even to
increase long before the plateau in $~\Pi~$ is reached.

Attempting to determine the reliability of the single precision CG
method in a more confidential way we compared the solutions for our
fermionic observables given by CGSP to those of CGDP applied to a set of
200 gauge configurations for various values of $~\kappa~$ in the
vicinity of $~\kappa_c~$ at $~\beta=0~$ on a $~8^3\times 16~$ lattice.
On every configuration $~k~$ we checked the relative deviation

\eq
\delta(O^k) = \frac{O^k_{DP}-O^k_{SP}}{O^k_{DP}}~,
                        \label{dev}
\en

\noi where $~O^k_{DP,SP}~$ denotes one of
$~\bar{\psi}\psi,~\bar{\psi}\gamma_5 \psi,~$ or $~\Pi~$ on the
particular configuration obtained by CGDP or CGSP, respectively.  Some
results are given in Figure {\ref{fig:3_1b}}.
In the case of $~\kappa=0.25~$ (see Figure {\ref{fig:3_1b}}a) the fermionic
observables seem to be quite well reproduced by CGSP.  The maximal relative
deviations show up in $~\bar{\psi}\gamma_5\psi$, but remain under 3\%.  
For the averages we obtain:  $~\langle \delta(\bar{\psi}\psi) \rangle =
-3\times10^{-5}\pm 4\times 10^{-4},~ \langle
\delta(\bar{\psi}\gamma_5\psi) \rangle = 2\times 10^{-4} \pm 3\times
10^{-3}, ~ \langle \delta(\Pi) \rangle = -9 \times 10^{-6} \pm 2\times
10^{-3}~.$ Therefore, the average 'IEEE--noise' coming from the
comparison of single to double precision data is well--consistent with
zero within errorbars.
However, at another -- somewhat lower -- value 
of $~\kappa~$ (Figure {\ref{fig:3_1b}}b) CGSP provided results 
which significantly differed from 
those of CGDP on some configurations and hence cannot be ignored.
In particular, the maximal $~|\delta(\pbp)|~$ is about $~\sim 20\%$.
For the averages of $~\delta~$ we obtain 
$~\langle \delta(\pbp) \rangle \sim  -0.002 \pm 0.02~,
~~\langle \delta(\bar{\psi}\gamma_5\psi) 
\rangle \sim -0.0004 \pm 0.007~$ and 
$~\langle \delta(\Pi)  \rangle \sim -0.0007 \pm 0.09$,
which, however, are still well consistent with zero.
The maximal relative deviations observed on other 
values of $~\kappa~$ have been $~\sim 6 \div 39\%$, demonstrating that 
CG in single precision mode is not {\it a priori} reliable 
in the 'critical' region. Although the net effect of single precision on 
the CG method seemed to be small, i.e. the found averages of $~\delta~$'s are 
compatible with zero, it might happen that the averages of observables
suffer from single precision, especially on sets of small statistics.

Therefore, the question about the reliability of the single precision
(even Kahan improved) CG still remains open in the case of 
the confinement phase. 
More detailed study, especially on larger lattices, is necessary
to draw a final conclusion here.

\begin{figure}[p]
\vspace{-3.0cm}
\epsfysize=570pt\epsfbox{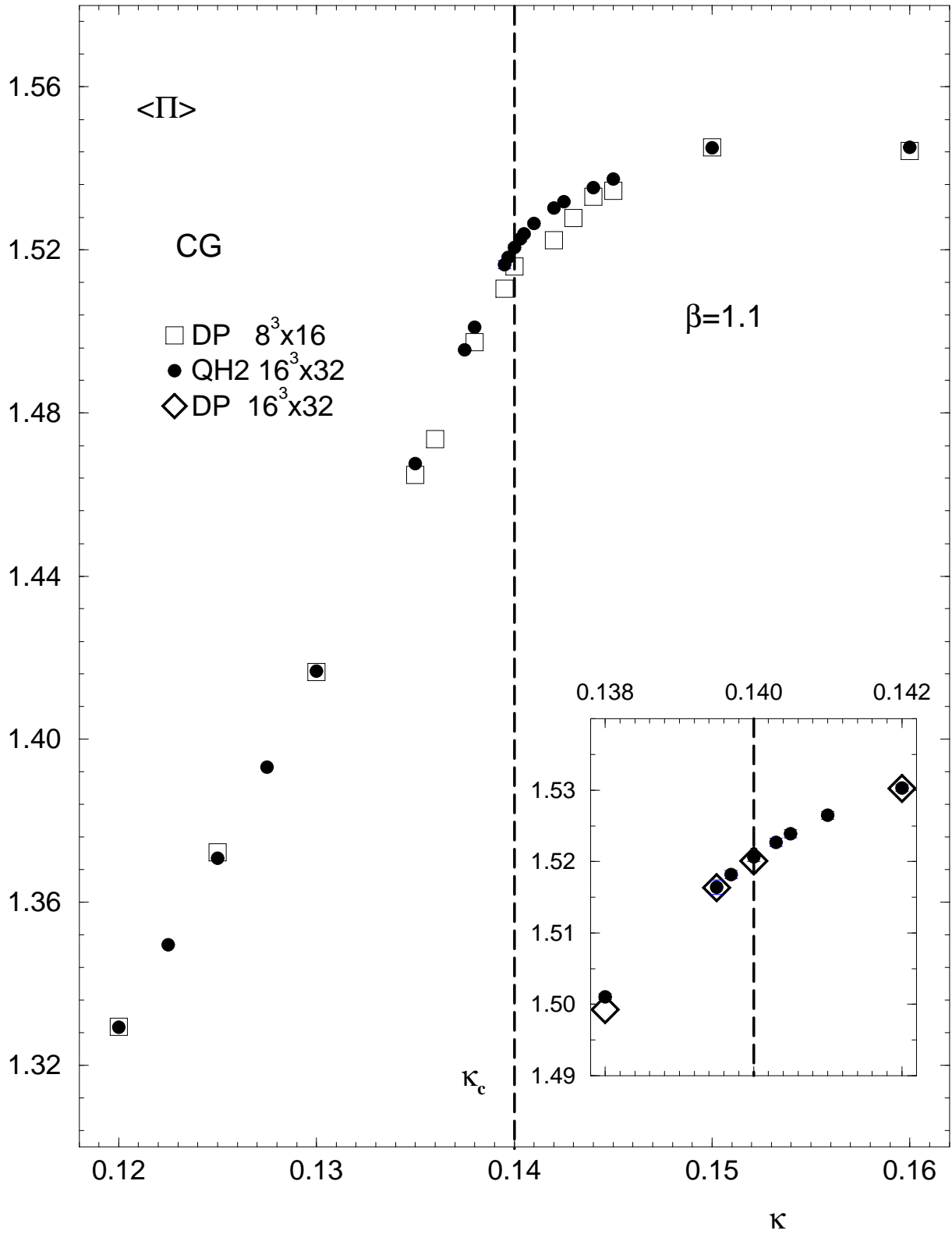}
\vspace{1.0cm}
\caption{
$~\langle \Pi \rangle~$ as a function of $~\kappa~$ at $~\beta=1.1$
obtained by CGSP (QH2) and CGDP.
The vertical dashed line indicates the position of $~\kappa_c~$. 
}
\label{fig:cgsp_coul}
\end{figure}

Fortunately, the situation is much more favorable in the 
Coulomb phase. The inversion procedure
is comparatively fast both for $~\kappa ~\aleq~ \kappa_c~$ and 
$~\kappa > \kappa_c$. For all our observables we found very reasonable
agreement between single precision and double precision calculations.
As an example, we show in Figure \ref{fig:cgsp_coul}
the average values of the pion norm $~\Pi~$ calculated with 
single (QH2) and double precision at $~\beta =1.1$.
Since most of the DP data are obtained on a smaller lattice than the
QH2 data there is a (rather weak) dependence 
of $~\langle \Pi \rangle~$ on the lattice size
in the 'critical' region 
around $~\kappa_c$. For the other values 
of $~\kappa~$ the agreement between CGDP and CG on the QH2 
is very good. 
Since the vicinity of $~\kappa_c~$ is most interesting, we calculated
four datapoints around $~\kappa_c~$ with CGDP on a $~16^3\times 32~$
lattice in order to compare them with the QH2 results from the 
same lattice size (see the 
inset of Figure \ref{fig:cgsp_coul}).
Obviously, the QH2 implementation of CG provides
correct results at this $~\bt~$ value.

%
%
%
%  fig_3_1c
%

\subsection{Double precision BiCG--stabilized methods}
%=================================================

\begin{figure}[p]
\begin{center}
\vspace{-3.0cm}
\leavevmode
\hbox{
\epsfysize=570pt\epsfbox{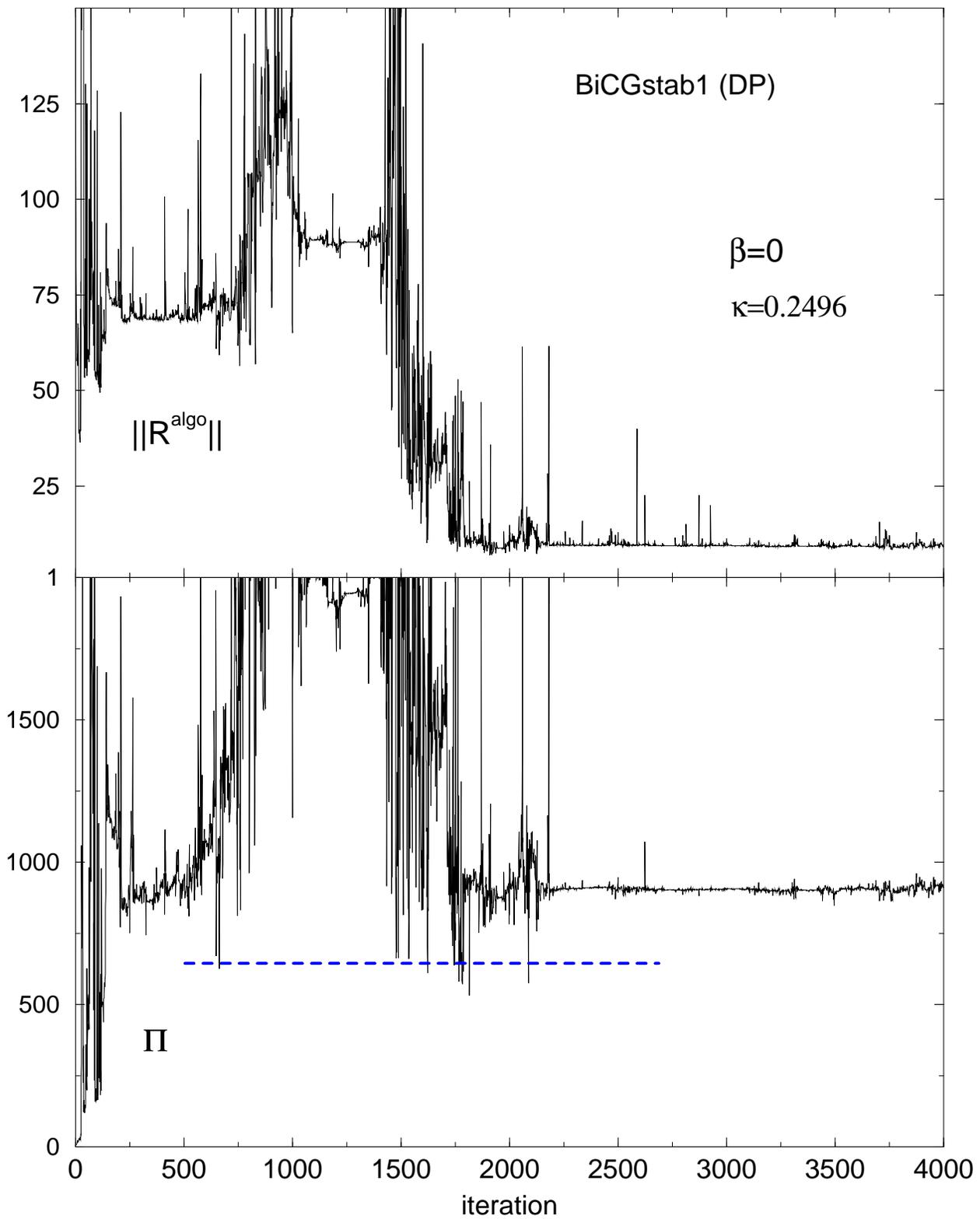}
     }
\end{center}
\vspace{1.0cm}
\caption{
An inversion history of BiCGstab1 in double precision mode 
at $\beta=0$ and $\kappa=0.2496$. 
The dashed horizontal line marks the CGDP solution for $~\Pi~$.
After $20000$ iterations the inversion process has been aborted.
}
\label{fig:Niter_b1_dp}
\end{figure}

As in the previous subsection for CG we compared the 'stabilized' methods 
by measuring
fermionic observables on the same configurations.  As long as the
hopping parameter $~\kappa~$ was not too close to $~\kappa_c(\beta)~$
all three methods agreed with very high precision for all
$~\beta$--values we used.  However, at $~\kappa \to \kappa_c(\beta )~$
failures occurred even in the double precision mode, i.e.  either the 
method did not converge at all or (as
in the case of BiCGstab2) provided a solution deviating significantly
from CGDP.  As an example of non--convergence of BiCGstab1 we show in Figure
\ref{fig:Niter_b1_dp} the first 4000 iterations of one inversion process
at $~\beta =0~$ and $~\kappa =0.2496$.
After $~20000~$ iterations we finally terminated the iterations.
The norm of the residue vector $~||R^{algo}||$
($||R^{true}||=||R^{algo}||$) did not decrease as expected and
always remained larger than one.
Both $~||R^{algo}||~$ and $~\Pi~$ stayed approximately constant
when the number of iterations became larger than $~2500$.
For comparison the dashed horizontal line shows the CGDP solution for 
the pion norm $~\Pi$.  
Restarting the procedure in some cases can help to resolve the problem.
However, the effect of restarting seems to be rather sensible to the
chosen frequency (see, e.g., \cite{templ}, p.22).

Tables \ref{tab:one} to \ref{tab:three} show values of $~\beta~$ and
$~\kappa~$ where failures of the different methods have been observed.
The convergence of BiCGstab1 demonstrates an interesting
$~\beta$--dependence.  While in the extreme strong coupling limit
$~\beta =0~$ it failed on some configurations in the limit $~\kappa \to
\kappa_c(\beta )$,  
we did not observe failures at the larger $~\beta$--values
in this limit.
Failure rates of BiCGstab1 will be discussed in the next section.

\begin{figure}[p]
\begin{center}
\vspace{-3.0cm}
\leavevmode
\epsfysize=600pt\epsfbox{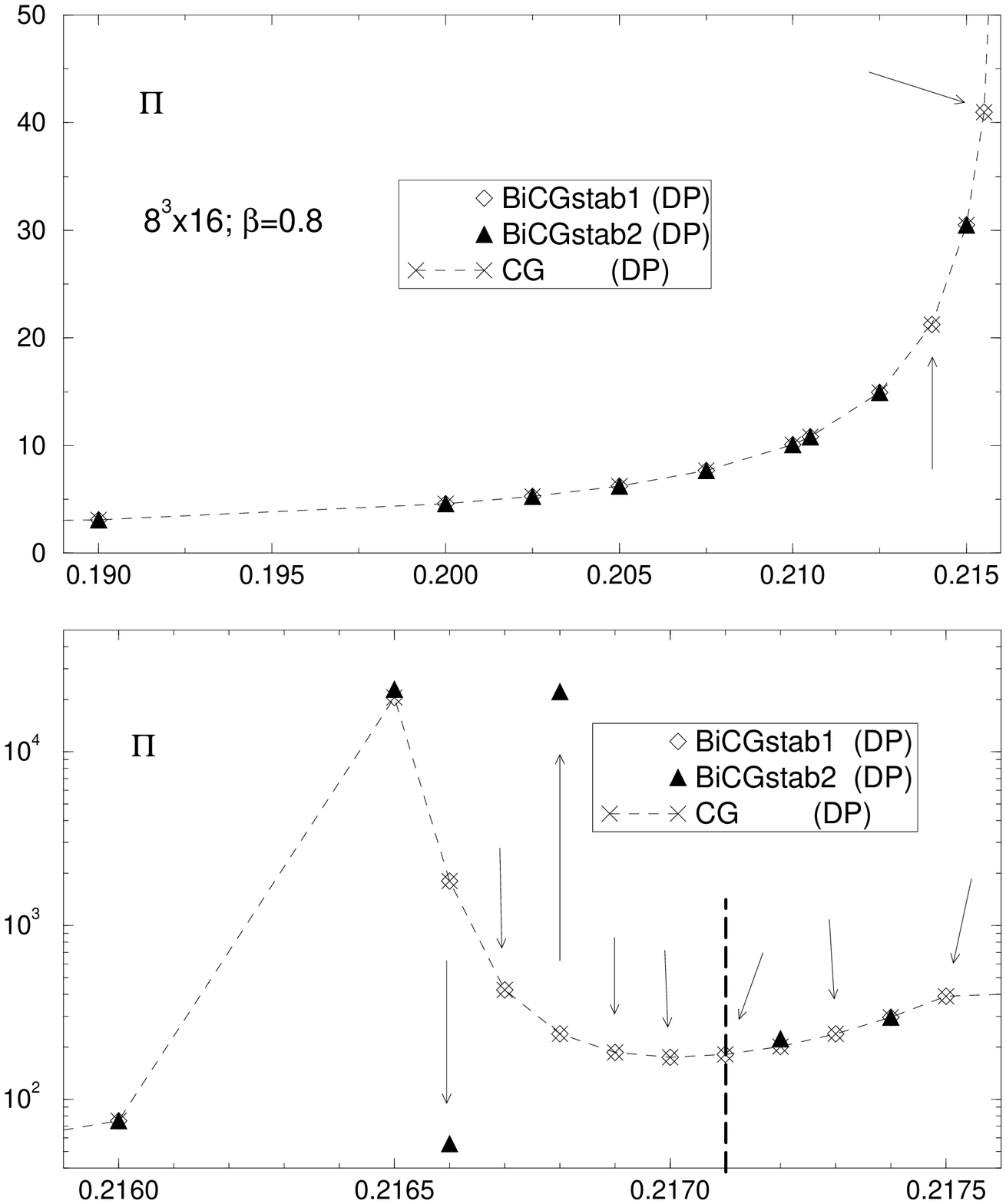}
\end{center}
%\vspace{1.0cm}
\caption{
Pion norm $~\Pi ~$ on individual configurations 
for CG, BiCGstab1 and BiCGstab2 in double precision mode. 
On the lower plot the $~\kappa~$ range is chosen to be in the 
vicinity of $~\kappa_c(\beta)~$ (vertical line). 
Arrows indicate where BiCGstab2 fails.
}
\label{fig:Pi_CX}
\end{figure}

As can be seen from these Tables BiCGstab2 appears to be less reliable
when $~\kappa \sim \kappa_c$.  Figure \ref{fig:Pi_CX} illustrates the
situation for different $~\kappa$--values at $~\beta =0.8$.  For every
$~\kappa$--value the pion norm $~\Pi~$ was calculated with the different
methods on the same configuration.  In the 'critical' region (lower plot
in Figure \ref{fig:Pi_CX}) BiCGstab1 matches CG with high
precision, while BiCGstab2 fails.  Near to $~\kappa_c(\bt)$ the number
of failures of BiCGstab2 turned out to be very large.  Therefore, we
conclude that BiCGstab2 even in double precision mode is unsuitable
for matrix inversion in the limit $~\lambda_{min} \to 0~$, i.e.,
$~\kappa \to \kappa_c(\beta )$.  The comparative complexity of this
algorithm turns out to be rather a disadvantage for the range of
investigated values of $~\bt~$ and $~\kappa$.  In particular, the
calculation of the optimal coefficients $~\xi_m~$ and $~\eta_m~$ (see
Section 2 and \cite{bcg2a}) can be the source of numerical instability,
since it involves operations like extra dot products, matrix inversion
and rather complicated recurrences where roundoff errors can
dramatically accumulate. 
This is quite likely what happens in
the case $~\kappa
\ra \kappa_c(\bt)$.

We also performed runs at different $~\beta$'s 
and $~\kappa$'s in the 'critical' region to compare the averages
of different fermionic observables.
Apart from the pion norm and fermionic condensate we
calculated also for every configuration k the pseudoscalar correlator

$$
\Gamma_k (\tau ) \sim \sum_{{\vec x},{\vec y}}
\left\{
\tr \Bigl( \cam^{-1}_{\, xy} \gamma_5\cam^{-1}_{\, yx} \gamma_5 \Bigr)
- \tr \Bigl( \cam^{-1}_{\, xx} \gamma_5 \Bigr) \cdot
\tr \Bigl( \cam^{-1}_{\, yy} \gamma_5 \Bigr) \right\}~,
\quad \tau =|x_4-y_4|
$$

\begin{figure}[p]
\begin{center}
\vspace{-3.0cm}
\leavevmode
\epsfysize=630pt\epsfbox{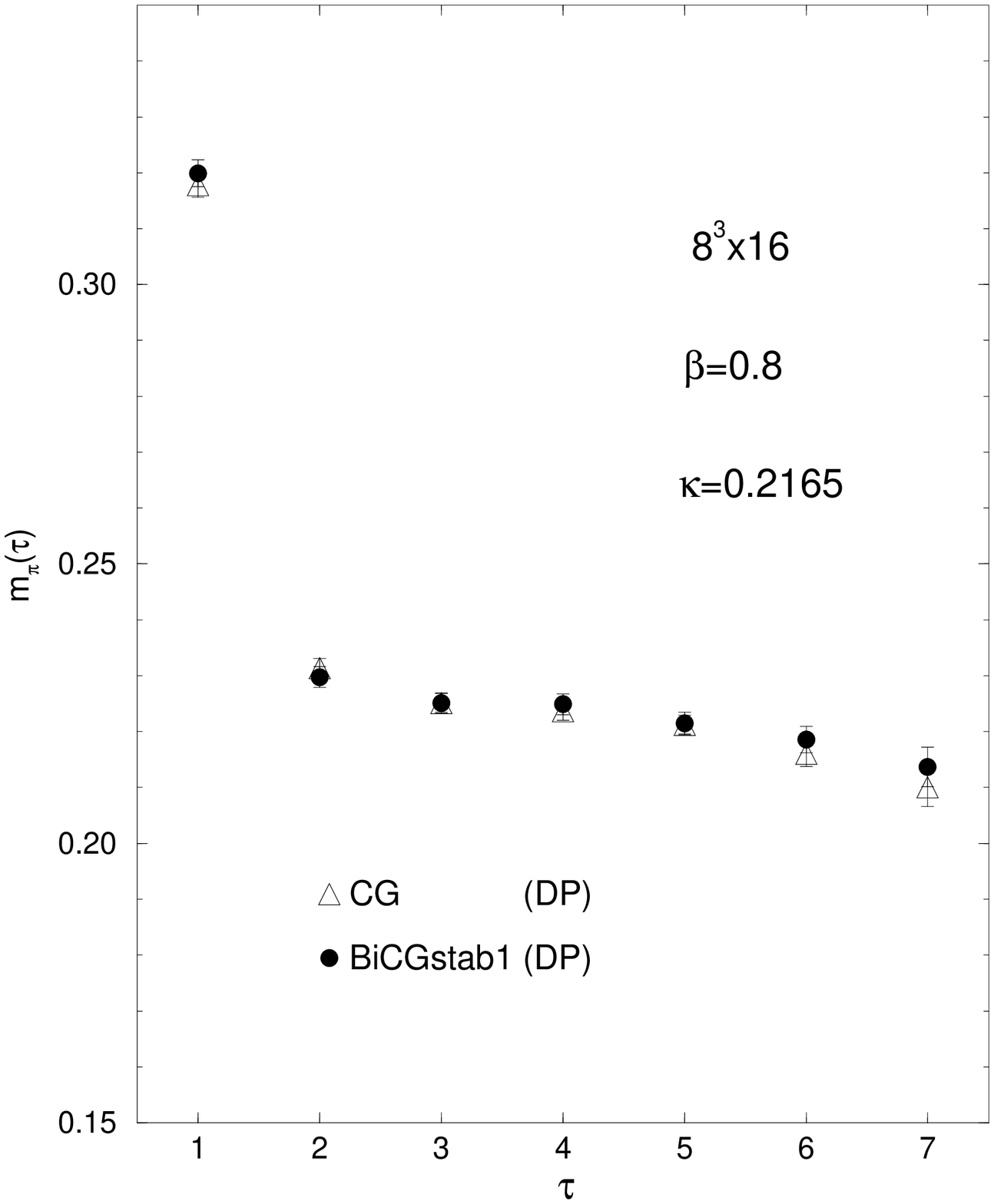}
\end{center}
%\vspace{1.0cm}
\caption{
 The $~\tau~$ dependence of the effective pseudoscalar mass $~m_{\pi}(\tau)$.
 The number of measurements is 
 $~\sim 8000$.
 } 
\label{fig:M_pi_CX}
\end{figure}

\noi to determine the effective pion mass $~m_{\pi}(\tau )$.
For $~\kappa$--values close to $~\kappa_c~$ we used an estimator defined
in \cite{hmm_95} to obtain a clear signal of $~m_{\pi}(\tau)$.  In
Figure \ref{fig:M_pi_CX} we show the $~\tau$--dependence of the
effective 'pion' mass $~m_{\pi}(\tau )~$ at $~\beta=0.8~$ and
$~\kappa=0.2165$.  Evidently, BiCGstab1 provides results well consistent
with that obtained by CG.

%
%  fig_3_2a
%
%
%

%
%  fig_3_2b
%
%
%

%
%  fig_3_2c
%

\subsection{Single precision BiCG--stabilized methods}
%=================================================

This section is devoted to the comparison of the 
single precision and
double precision BiCGstab methods.
The main accent was made on the TAO--language 
realization of these algorithms on the APE/Quadrics machine (QH2).
Since BiCGstab2 turned out to be
unreliable (and in addition less efficient than 
BiCGstab1 -- see the next section) for $~\kappa~$'s close 
to $~\kappa_c(\beta)~$
even in double precision we excluded this algorithm from
further study on the QH2.

\begin{figure}[p]
\begin{center}
\vspace{-3.0cm}
\leavevmode
\epsfysize=570pt\epsfbox{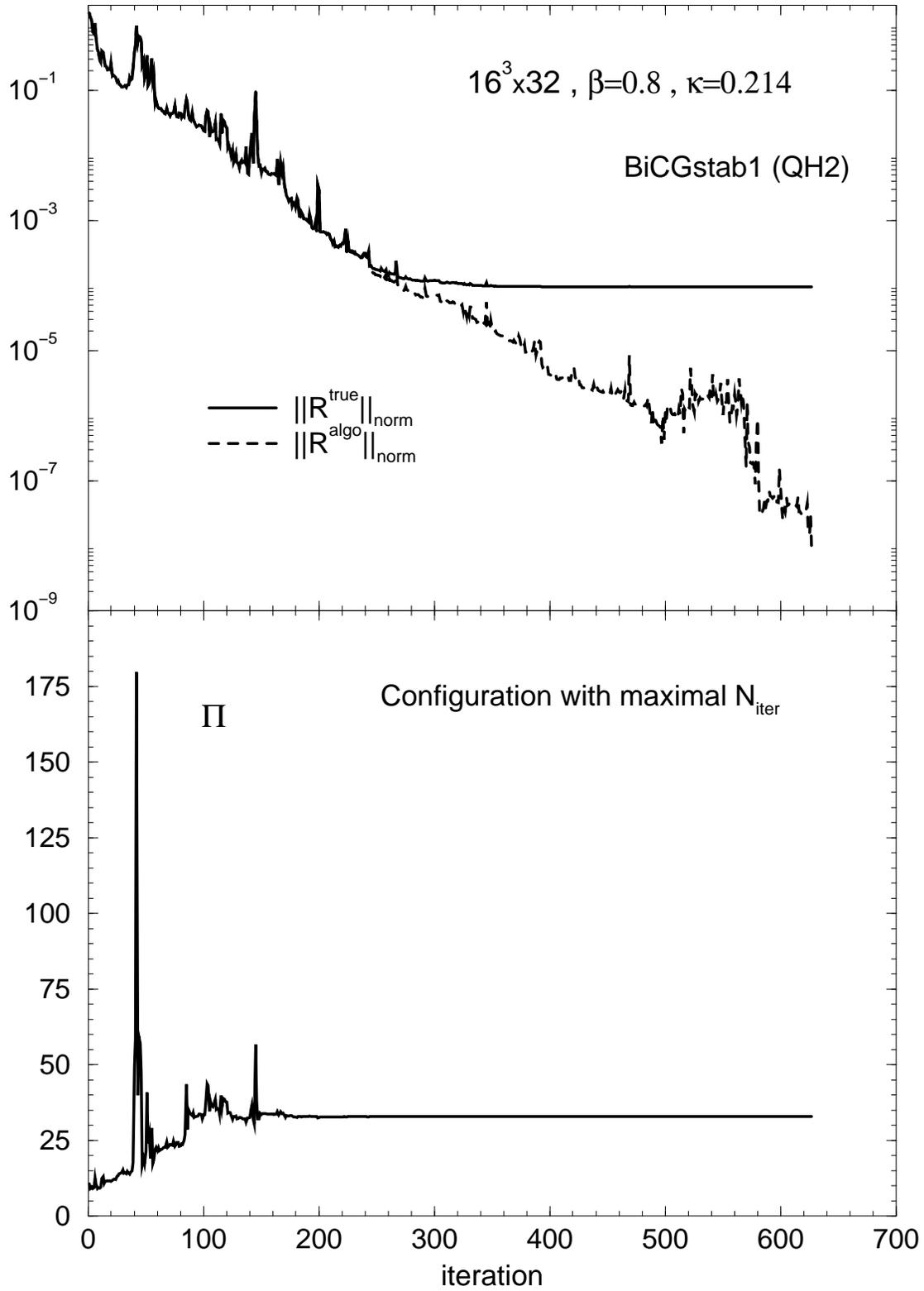}
\end{center}
\vspace{1.0cm}
\caption{
Normalized $~||R^{algo}||~$ and $~||R^{true}||~$ vs. inversion iterations
for the BiCGstab1 method on the QH2 and 
the corresponding evolution of $~\Pi~$.
 }
\label{fig:R_stab1_QH2}
\end{figure}

As an illustration, we show in 
Figure \ref{fig:R_stab1_QH2} the inversion history
of the BiCGstab1--run performed on the QH2 at $~\beta=0.8$ 
and $~\kappa=0.214$.
We have chosen here the configuration with the maximal number
of iteration steps.
The accumulation of roundoff errors 
entails the ramification of the  two residues
$|| R^{algo}||$ and $||R^{true}||$.
This is similar to the case of CG in single precision
(compare to Figure \ref{fig:R_CX_SP_QH2}).
Note, that the Kahan summation which we used
does not substantially increase the accuracy 
compared to unimproved single precision.
%
%
%  fig_3_3a
%

\begin{figure}[p]
\begin{center}
%\vspace{-3.0cm}
\vspace{-4.0cm}
\leavevmode
\epsfysize=630pt\epsfbox{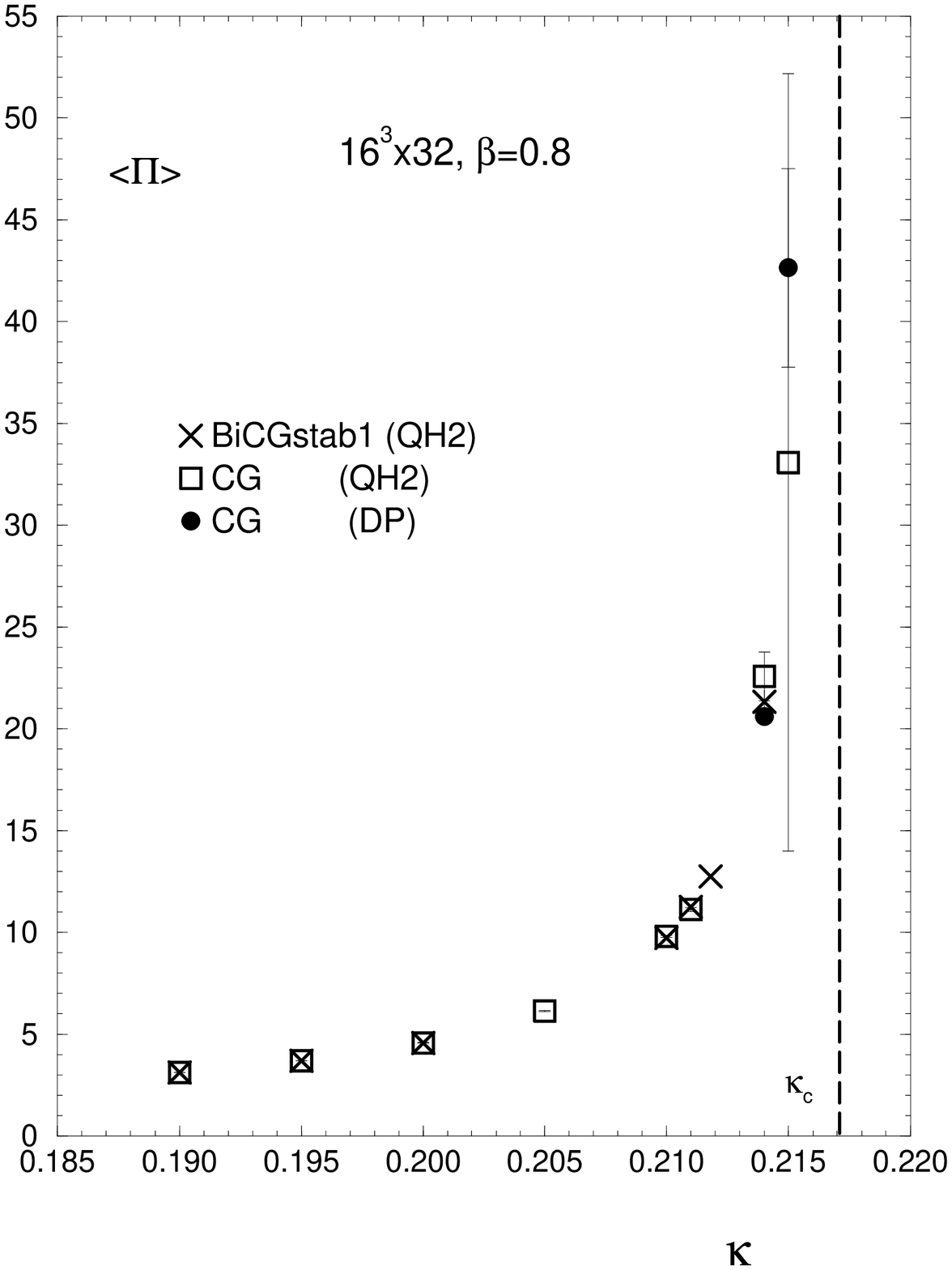}
\end{center}
\caption{
The average pion norm $~\langle \Pi \rangle ~$ as a function of 
$~\kappa~$ obtained from CG and BiCGstab1 on the QH2.
At $~\kappa=0.214;~0.215~$ we added also datapoints from CGDP.
The position of $~\kappa_c(\beta)~$ is indicated by the vertical 
dashed line.
 }
\label{fig:Pi_QH2}
\end{figure}

In Figure \ref{fig:Pi_QH2} we show the average of the pion norm 
$~\langle \Pi \rangle ~$ as a function of $~\kappa ~$ 
at $~\beta =0.8~$ on a $~16^3\times 32~$ lattice.
For $~\kappa < 0.214~$ we observe a very good agreement 
between BiCGstab1 and CG. We checked that these data are 
also well consistent with  CGDP data,  which
are not shown in this $~\kappa$--region.
%
%  fig_3_3b
%
For $~\kappa > 0.214~$ 
BiCGstab1 with single precision 
did not converge, in contrast to the double precision case at this 
value of $~\beta$.
For all inversion methods, statistical averages and errorbars
become ill-defined  due to 'exceptional' configurations
close to $~\kappa_c(\bt)$ \cite{hmm_95}.
This effect is observable already at $~\kappa=0.214~$
and especially at $~\kappa=0.215~$.

\setlength{\arrayrulewidth}{0.28mm}
\setlength{\doublerulesep}{1.5mm}
\renewcommand{\arraystretch}{1.3}
%   Table 1 
\begin{table}[htbp]
\centering
\begin{tabular}{|c|c|c|c|c|c|c|}\hline 
\multicolumn{7}{|c|}{\bf Confinement phase $~\beta=0.8~$}\\ \hline \hline
\multicolumn{2}{|c|}{ } & \multicolumn{2}{|c|}{\bf QH2} 
& \multicolumn{3}{|c|}{\bf Double Precision} \\ \hline 
 $\kappa$ & $m_q$ & CG & BiCGs1 
& CG & BiCGs1 & BiCGs2 \\ \hline \hline
         0.1900 & 0.3285 &$\surd$ & $\surd$
                          & $\surd$ & $\surd$ & $\surd$ \\ \hline
         0.1950 & 0.2610 &$\surd$ & $\surd$
                          & $\surd$ & $\surd$ & $\surd$ \\ \hline
         0.2000 & 0.1969 &$\surd$ & $\surd$
                          & $\surd$ & $\surd$ & $\surd$ \\ \hline
         0.2100 & 0.0779 &$\surd$ & $\surd$
                          & $\surd$ & $\surd$ & $\surd$ \\ \hline
         0.2105 & 0.0722 &$\surd$ & $\surd$
                          & $\surd$ & $\surd$ & $\surd$ \\ \hline
         0.2114 & 0.0621 &$\surd$ & $\surd$
                          & $\surd$ & $\surd$ & $\surd$ \\ \hline
         0.2118 & 0.0576 &$\surd$ & $\surd$
                          & $\surd$ & $\surd$ & $\surd$ \\ \hline
         0.2125 & 0.0499 &$\surd$ & $\surd$
                          & $\surd$ & $\surd$ & $\surd$ \\ \hline
         0.2140 & 0.0334 &$\surd$ & $\surd$
                          & $\surd$ & $\surd$ & $\bigcirc$ \\ \hline
         0.2150 & 0.0225 &? & $\bigcirc$
                          & $\surd$ & $\surd$ & $\surd$ \\ \hline
         0.2160 & 0.0117 &?       & $\bigcirc$  
                          & $\surd$ & $\surd$ & $\bigcirc$ \\ \hline
         0.2170 & 0.0011 & ?      & $\bigcirc$ 
                          & $\surd$ & $\surd$ & $\bigcirc$ \\ \hline
         {\bf 0.2171} &  0.0000 &        &         
                          & $\surd$ & $\surd$ & $\bigcirc$ \\ \hline
         0.2172 &        &        &         
                          & $\surd$ & $\surd$ & $\bigcirc$ \\ \hline
         0.2173 &        &        &         
                          &$\surd$  & $\surd$ & $\bigcirc$ \\ \hline
\end{tabular}
\caption{\label{tab:one}
Reliability of inverters on  $~16^3 \times 32~$ 
(QH2) and $~8^3\times 16~$ lattices (double precision machine).
Symbol $~\surd~$ denotes correct fermionic observables; $~\bigcirc~$
indicates a failure of the method. A question mark means that
no conclusion could be drawn.  $\kappa_c(\beta )$ is given by boldface.
}
\end{table}

Tables \ref{tab:one}  to \ref{tab:three} 
summarize our results concerning the reliability of the inverter 
methods on the QH2 and on the 
double precision machines in the confinement 
and Coulomb phases.
In the confinement phase single precision BiCGstab1 and BiCGstab2 
were observed to fail at smaller
values of $~\kappa~$ compared to the runs with double
precision arithmetic, and therefore are unsuitable to 
explore the 'critical' region.
On the other hand, in 
the Coulomb phase BiCGstab1 on the QH2 proved to work
even for $~\kappa=\kappa_c(\bt)$.

\begin{table}[htb]
\centering
\begin{tabular}{|c|c|c|c|c|}\hline 
\multicolumn{5}{|c|}{\bf Confinement phase $~\beta=0~$}\\ \hline \hline
\multicolumn{2}{|c|}{ } 
& \multicolumn{3}{|c|}{\bf Double Precision} \\ \hline 
 $\kappa$ & $m_q$ 
& CG & BiCGs1 & BiCGs2 \\ \hline \hline

         0.2425 & 0.0619
                          & $\surd$ & $\surd$ & $\surd$ \\ \hline
         0.2450 & 0.0408
                          & $\surd$ & $\surd$ & $\surd$ \\ \hline
         0.2475 & 0.0202
                          & $\surd$ & $\surd$ & $\surd$ \\ \hline
         0.2480 & 0.0161
                          & $\surd$ & $\surd$ & $\surd$ \\ \hline
         0.2482 & 0.0145
                          & $\surd$ & $\surd$ & $\surd$ \\ \hline
         0.2484 & 0.0129
                          & $\surd$ & $\surd$ & $\surd$ \\ \hline
         0.2486 & 0.0113
                          & $\surd$ & $\surd$ & $\surd$ \\ \hline
         0.2488 & 0.0096
                          & $\surd$ & $\surd$ & $\surd$ \\ \hline
         0.2490 & 0.0080
                          & $\surd$ & $\surd$ & $\bigcirc$ \\ \hline
         0.2492 & 0.0064
                          & $\surd$ & $\surd$ & $\bigcirc$ \\ \hline
         0.2494 & 0.0048
                          & $\surd$ & $\bigcirc$ & $\bigcirc$ \\ \hline
         0.2496 & 0.0032
                          & $\surd$ & $\bigcirc$ & $\bigcirc$ \\ \hline
         0.2498 & 0.0016
                          & $\surd$ & $\bigcirc$ & $\bigcirc$ \\ \hline
         0.2499 & 0.0008
                          & $\surd$ & $\bigcirc$ & $\bigcirc$ \\ \hline
         {\bf 0.2500} & 0.0000
                          & $\surd$ & $\bigcirc$ &         \\ \hline
\end{tabular}
\caption{\label{tab:two}
Same as Table \protect\ref{tab:one} but at $~\beta=0$.
}
\end{table}

\begin{table}[htb]
\centering
\begin{tabular}{|c|c|c|c|c|c|c|}\hline 
\multicolumn{7}{|c|}{\bf Coulomb phase $~\beta=1.1~$}\\ \hline \hline
\multicolumn{2}{|c|}{ } & \multicolumn{2}{|c|}{\bf QH2} 
& \multicolumn{3}{|c|}{\bf Double Precision} \\ \hline 
 $\kappa$ & $m_q$ & CG & BiCGs1 
& CG & BiCGs1 & BiCGs2 \\ \hline \hline
%
%
%         0.13000 & 0.2747 & $\surd$&        
%                          & $\surd$ & $\surd$ & $\surd$ \\ \hline
%         0.13200 & 0.2165 &        &        
%                          & $\surd$ & $\surd$ & $\surd$ \\ \hline
%         0.13400 & 0.1599 &        &        
%                          & $\surd$ & $\surd$ & $\surd$ \\ \hline
         0.13500 & 0.1323 & $\surd$& $\surd$ 
                          & $\surd$ & $\surd$ &         \\ \hline
         0.13600 & 0.1050 &        & $\surd$
                          & $\surd$ & $\surd$ & $\surd$ \\ \hline
         0.13700 & 0.0782 & $\surd$&$\surd$ 
                          & $\surd$ & $\surd$ & $\surd$ \\ \hline
         0.13800 & 0.0518 & $\surd$& $\surd$
                          & $\surd$ & $\surd$ & $\surd$ \\ \hline
         0.13900 & 0.0257 & $\surd$&$\surd$ 
                          & $\surd$ & $\surd$ & $\surd$ \\ \hline
         0.13950 & 0.0128 & $\surd$ &$\surd$ 
                          & $\surd$ & $\surd$ &$\bigcirc$\\ \hline
         {\bf 0.14000} & 0.0000 &$\surd$ & $\surd$ 
                          & $\surd$ & $\surd$ & $\bigcirc$ \\ \hline
         0.14200 &        &$\surd$ & $\bigcirc$ 
                          & $\surd$ & $\surd$ & $\bigcirc$ \\ \hline
         0.14400 &        &$\surd$ & $\bigcirc$ 
                          & $\surd$ & $\surd$ & $\bigcirc$ \\ \hline
\end{tabular}
\caption{\label{tab:three}
Same as Tables \protect\ref{tab:one}, \protect\ref{tab:two}
but in the Coulomb phase at $~\beta=1.1$.
}
\end{table}

\begin{figure}[p]
\begin{center}
\vspace{-3.0cm}
\leavevmode
\epsfysize=630pt\epsfbox{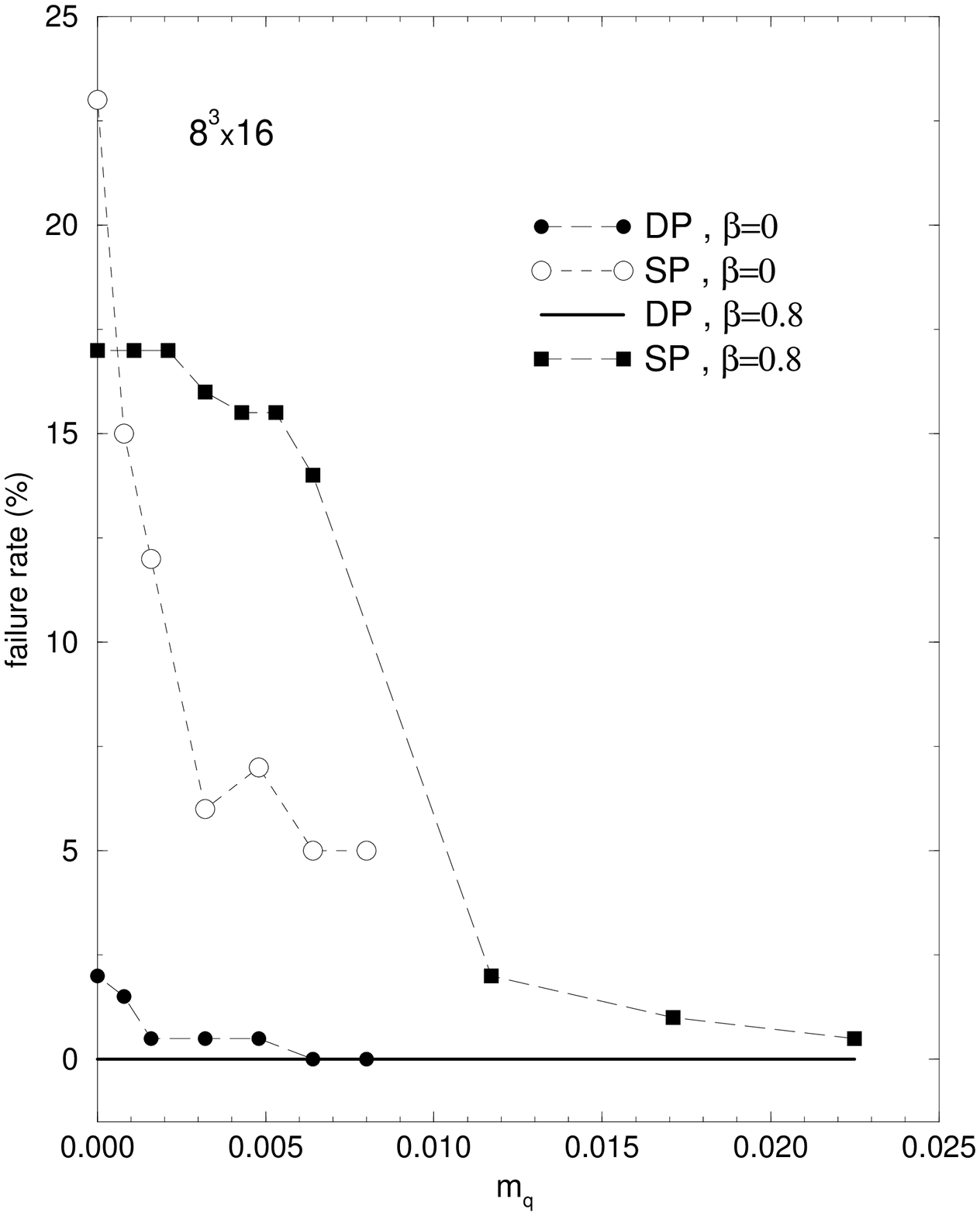}
\end{center}
%\vspace{1.0cm}
\caption{
The failure rate (in $\%$) of 
BiCGstab1 for $~\beta=0~$ and 
$~\beta=0.8~$ in double and single precision modes.
$m_q~$ is defined in eq.(\protect\ref{mq}).
 }
\label{fig:3_3c}
\end{figure}

To quantify the results of Tables \ref{tab:one}, \ref{tab:two}
we show in Figure \ref{fig:3_3c} the failure rates of the
single and double precision BiCGstab1 implementations
in dependence on $~m_q$ obtained from 210 gauge configurations.
We allowed the fermionic observables to deviate by a certain
cutoff from the corresponding CGDP result
on each configuration. The failure rates turned out to be 
almost insensitive  to the special choice 
of the cutoff in both, single and double 
precision cases, when it was varied from $0.5\%$ to $40\%$.
The results confirm, that the failure rate is drastically 
increased when single precision is used. It suggests also, that the 
double precision version of BiCGstab1 may be applied 
very close to $~\kappa_c~$ in the confinement phase. 
For instance, at $~\beta=0.8~$ the failure rate of the double precision
BiCGstab1 has been found to be zero for all investigated
values of $~\kappa$.

%
%  fig_3_c
%

\section{Efficiency}
%======================

\begin{figure}[p]
\begin{center}
\vspace{-3.0cm}
\leavevmode
\epsfysize=570pt\epsfbox{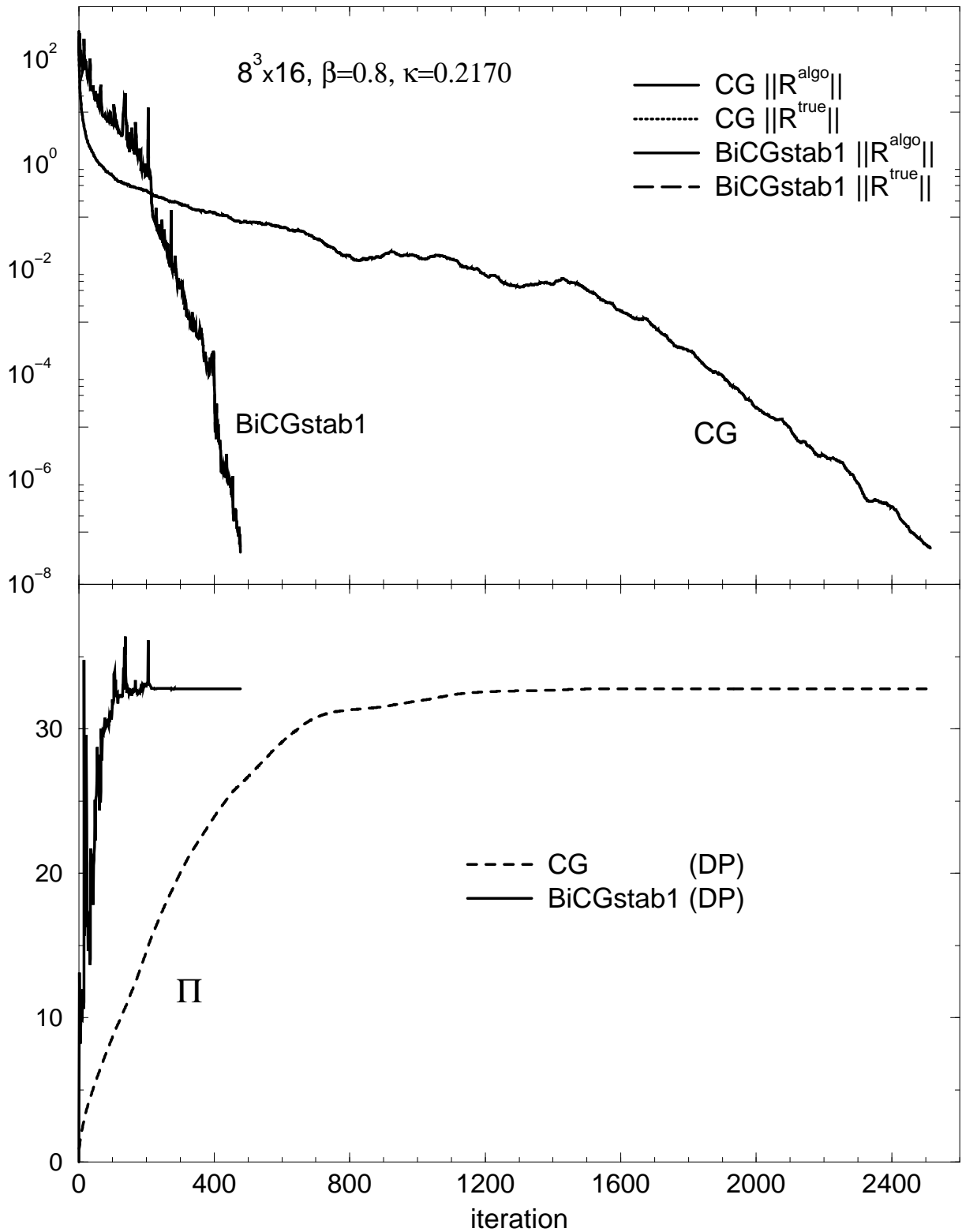}
\end{center}
\vspace{1.0cm}
\caption{
Convergence history for the double precision CG and BiCGstab1
at $\beta=0.8$ and $\kappa =0.2170$.
 }
\label{fig:inv_hist}
\end{figure}

In the confinement phase the double precision BiCGstab1 demonstrates
a pronounced superiority over CG and BiCGstab2 up to values of
$~\kappa~$ rather close to $~\kappa_c(\beta )$. As an example, we show
in Figure \ref{fig:inv_hist} the inversion history for CG and BiCGstab1
at $~\beta=0.8$ and $~\kappa =0.2170$ ($\kappa_c = 0.2171(1)$)
on an $~8^3 \times 16~$ lattice 
in double precision mode.
Both methods were applied to the same configuration.
The data points for $~||R^{algo}||~$ and $~||R^{true}||~$ are on top of each
other for both methods. The pion norm $~\Pi~$  reaches a plateau before 
the stopping criterion is fulfilled.
As far as CG and BiCGstab1 require the same number of matrix multiplications
per iteration their CPU time requirement per iteration is about the same.
Viewed in this context, the small number of BiCGstab1 iterations compared to
CG is impressive.

\begin{figure}[p]
\begin{center}
\vspace{-3.0cm}
\leavevmode
\epsfysize=630pt\epsfbox{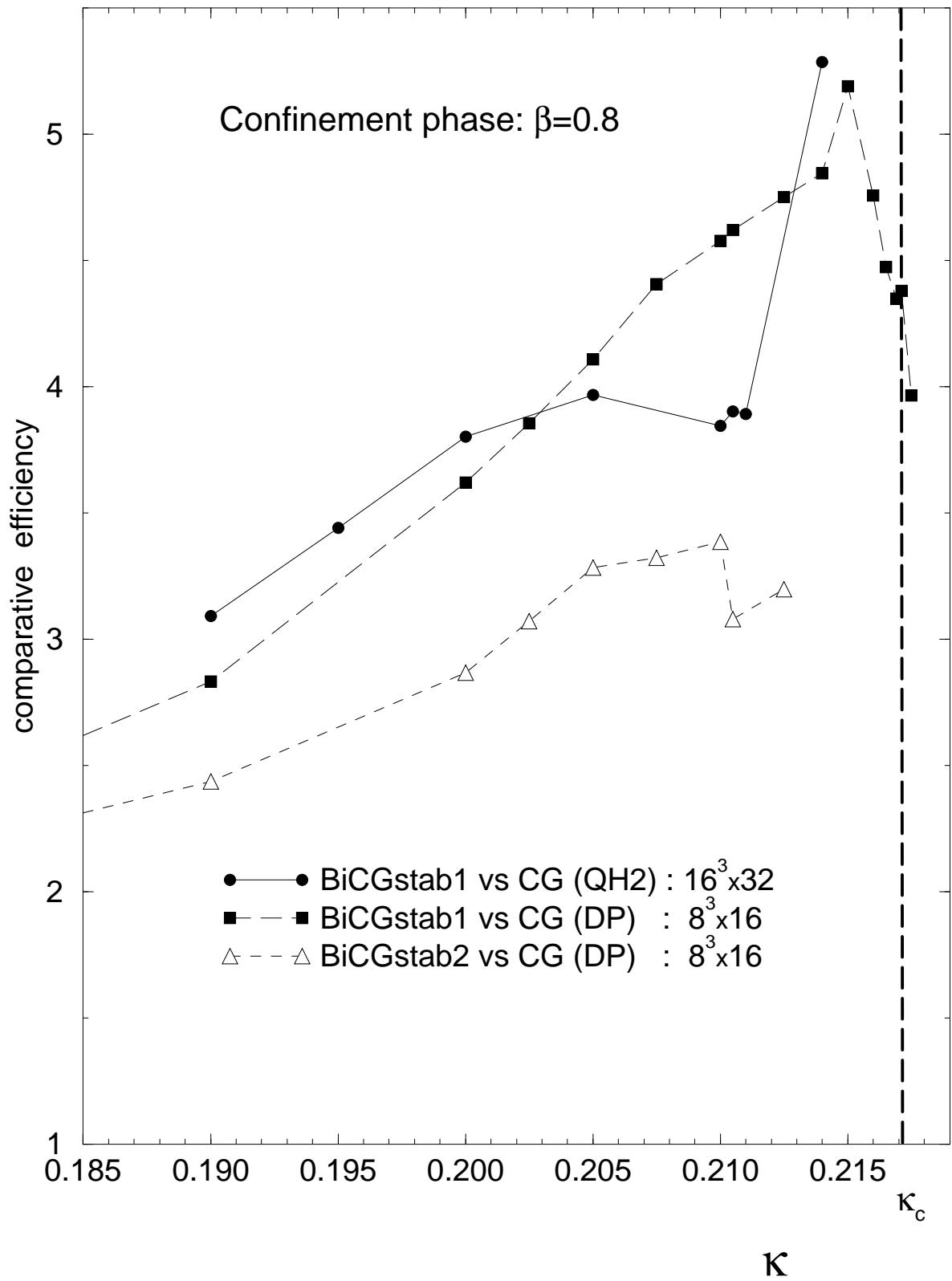}
\end{center}
\vspace{1.0cm}
\caption{
Comparative efficiency of BiCGstab1 and BiCGstab2
for different $\kappa$'s in the 
confinement phase at $\beta=0.8$. The vertical line indicates
the position of $~\kappa_c$.
 }
\label{fig:Effic_08}
\end{figure}

To quantify this point we compared the average CPU--times $~t_{cpu}~$ per
inversion between the different methods. 
In the case of BiCGstab2 we compared $~t_{cpu}~$ for 10 configurations
to the CPU--time  required by BiCGstab1 on the {\it same} set of
configurations including the same initial conditions. It turned out, that
in the confinement phase at $~\bt=0.8~$ BiCGstab2 is by a factor 
$~\sim 0.6 \div 0.9~$ less efficient than BiCGstab1, depending on the 
choice of $~\kappa$. 
In Figure \ref{fig:Effic_08} we display the 
comparative efficiency, i.e. 
the ratios $~t_{cpu}^{CG}\big/t_{cpu}^{BiCGstab1}~$ and
$~t_{cpu}^{CG}\big/t_{cpu}^{BiCGstab2}~$
at $~\bt=0.8$ in dependence on $~\kappa$.
The number of measurements for 
BiCGstab1 on the double precision machine has been 200, while on the 
QH2 it ranged from $\sim80\div800$.

BiCGstab1 yields a 
considerable gain in CPU time by a factor of $~\sim 3 \div 5 ~$ on the QH2.
This factor is supported by the double precision data and
is comparable with findings from quenched QCD tests \cite{bcg1b}.
Note, that the comparative efficiency is rather stable with 
respect to the lattice
size, since Figure \ref{fig:Effic_08} displays 
data from $~8^3\times 16~$ and $~16^3\times 32~$ lattices.
In the case of $~\bt=0~$ the same 
picture holds qualitatively, however the maximal CPU time improvement 
factor was around 9 on double precision vector machines. 

As mentioned, BiCGstab2, if reliable, was always less efficient than 
BiCGstab1 but a factor of $~\sim 2 \div 3~$ better than CG.

\begin{figure}[p]
\begin{center}
\vspace{-3.0cm}
\leavevmode
\epsfysize=625pt\epsfbox{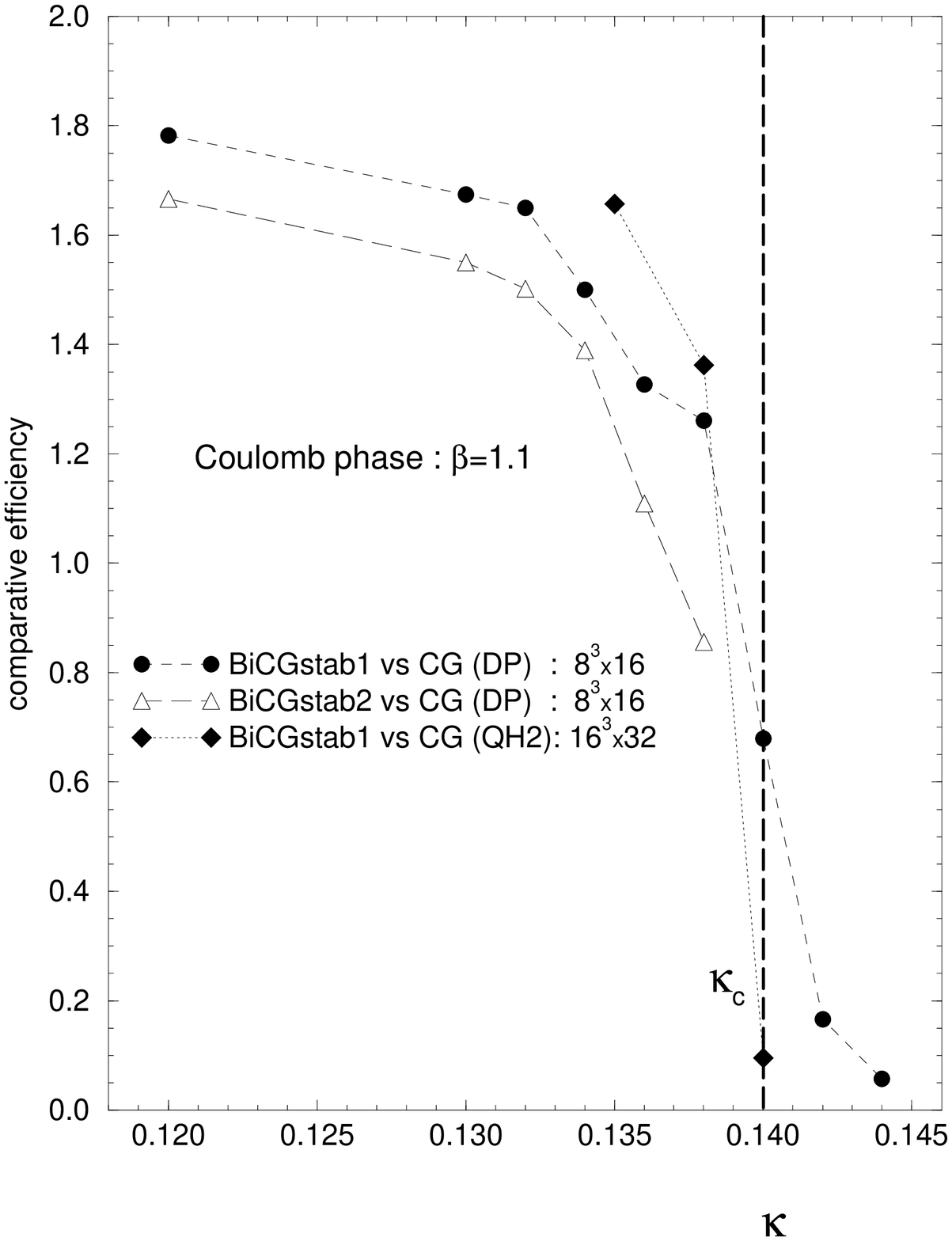}
\end{center}
\vspace{1.0cm}
\caption{
Same as Figure \protect\ref{fig:Effic_08} but
in the Coulomb phase at $\beta =1.1$. 
 }
\label{fig:Effic_11}
\end{figure}

Things change dramatically in the Coulomb
phase (see Figure \ref{fig:Effic_11}).
The comparative efficiency of BiCGstab1 displays a sudden 
drop around $~\kappa_c$, the trend for BiCGstab2 is the same.
Sufficiently below $~\kappa_c~$ the maximal
factor of CPU time improvement is below 2 and
goes to zero for $~\kappa \rightarrow \kappa_c(\beta)~$.
BiCGstab2 has always found to be less
efficient than BiCGstab1. 

Presumably, the completely different distributions of eigenvalues
of the fermionic matrix in the confinement and Coulomb phases
cause this significantly different behavior of the
inversion algorithms.

%
%
%  fig_4_1a
%
%
%

%
%
%  fig_4_1b
%
%
%

%
%
%  fig_4c
%

\section{Summary  and conclusions}
%=================================

To summarize, we studied the reliability and comparative efficiency
of the three matrix inversion methods : CG, BiCGstab1 and
BiCGstab2. 
For this study we employed the $~U(1)~$ lattice gauge 
theory with Wilson fermions, the inverted matrix
being 
an even-odd decomposed version of the  
Wilson fermionic matrix $~{\cal M}~$ 
(or $~{\cal M}^{\dagger}{\cal M}$).
The main accent in our study was made on the 'critical'
case $\kappa \sim \kappa_c(\beta)$.
The runs have been performed in double precision 
implementation (Cray, Convex, Siemens--Fujitsu) and in 
single precision implementation (Convex and especially QH2).

Our main conclusions are as follows.

\begin{itemize}

\item The standard CG with double precision appears to be the most
reliable method, though not the fastest one. No failures were
observed even at $~\kappa ~\ageq~ \kappa_c(\beta )$.

The CG with single precision in the Coulomb phase,
i.e., $\beta > \beta_0 \simeq 1.01~$, is very reliable and reproduces 
the double precision results with reasonable accuracy even
at $~\kappa =\kappa_c(\beta )$.

In the confinement phase ($\beta < \beta_0 $), however, the question
of the reliabily of the CGSP in the $~\kappa \to \kappa_c(\bt)~$ limit
still remains open.

\item  The reliability of the double precision BiCGstab1 is 
rather high
when $~\kappa \sim \kappa_c(\beta )$. 
The failure rate is zero at $~\beta =0.8~$, and 
about 2\% in the extreme case $~\beta =0$. 
Given some precautions for a fatal case, e.g. restarting 
or switching to another inverter, it 
can be used to study the limit $~\kappa \ra \kappa_c(\bt)$.

In the case of single precision in the confinement phase
the reliability of the BiCGstab1 drastically drops in the
$~\kappa \to \kappa_c(\beta )~$ limit.

\item In the confinement phase in the $~\kappa$--region, where a comparison 
was possible, BiCGstab1 turned out to be the most efficient algorithm.
The gain in CPU time was a factor of $~\sim 2.5 \div 5~$  as compared to CG 
at $~\beta=0.8$, and up to a factor of $~\sim 9$ in the extreme 
strong coupling regime $~\beta=0$.

This does not hold for the Coulomb phase, where BiCGstab1 
(as well as BiCGstab2) exhibits a sudden drop in efficiency around
$~\kappa_c(\beta)~$ and CG becomes the fastest method. 

\item BiCGstab2 even with double precision shows rather low
reliability in the 'critical' region $~\kappa \sim \kappa_c$.
This fact
practically excludes this method from the study of the
chiral limit of the theory.

In the parameter range, where BiCGstab2 worked, it was always 
less efficient than BiCGstab1 (by a factor $~\sim 0.6 \div 0.9$).

\end{itemize}

\noi 

The completely different behavior of the comparative efficiency found in
the confinement and Coulomb phases suggests, that the optimal choice of
the inversion algorithm is dictated by the typical distribution of
eigenvalues in the particular phase.  Further investigation is needed to
work out this point.

We expect that most of the conclusions drawn in this paper
can be applied also to lattice QCD.

%\vspace{0.5cm}

\pagebreak

\noi {\large {\bf Acknowledgements}}

A.H. and V.K.M acknowledge support by the Deutsche
Forschungsgemeinschaft with research grant Mu 932/1-4.  
The calculations were performed in DESY--IFH Zeuthen, Konrad--Zuse--Zentrum
Berlin, RRZN Hannover and RZ at Humboldt University Berlin.
A.H., V.K.M. and M.M.-P. are grateful to Fr. Subklev from RZ for support.

%\pagebreak

\end{document}